\documentclass[twocolumn,showpacs,preprintnumbers,prb]{revtex4-2}

\usepackage{graphicx,color}

\usepackage{dcolumn}

\usepackage{bm}

\usepackage[caption=false]{subfig}
\begin{document}

\preprint{}

\title{Magnetic excitations from the hexagonal spin clusters in the $S$ = 1/2 distorted honeycomb lattice antiferromagnet Cu$_2$(pymca)$_3$(ClO$_4$)}

\author{Masaaki Matsuda}

\affiliation{Neutron Scattering Division, Oak Ridge National Laboratory, Oak Ridge, Tennessee 37831, USA}

\author{Alexander I. Kolesnikov}

\affiliation{Neutron Scattering Division, Oak Ridge National Laboratory, Oak Ridge, Tennessee 37831, USA}

\author{Zentaro Honda}

\affiliation{Graduate School of Science and Engineering, Saitama University, Sakura-ku, Saitama 338-8570, Japan}

\author{Tokuro Shimokawa}

\affiliation{Theory of Quantum Matter Unit, Okinawa Institute of Science and Technology Graduate University, Onna 904-0495, Japan}

\author{Sho Inoue}

\affiliation{Center for Advanced High Magnetic Field Science, Graduate School of Science, The University of Osaka, Toyonaka, Osaka 560-0043, Japan}

\author{Yasuo Narumi}

\affiliation{Center for Advanced High Magnetic Field Science, Graduate School of Science, The University of Osaka, Toyonaka, Osaka 560-0043, Japan}

\author{Masayuki Hagiwara}

\affiliation{Center for Advanced High Magnetic Field Science, Graduate School of Science, The University of Osaka, Toyonaka, Osaka 560-0043, Japan}

\date{\today}

\begin{abstract}

Cu$_2$(pymca)$_3$(ClO$_4$) (pymca: pyrimidine-2-carboxylate) consists of a slightly distorted honeycomb lattice of Cu$^{2+}$ spins, which shows no long-range magnetic order down to 0.6 K. A magnetization study revealed 1/3 and 2/3 plateau phases
[A. Okutani $et$ $al.$, J. Phys. Soc. Jpn. {\bf 88}, 013703 (2019)], which is not expected for regular honeycomb antiferromagnets.
Inelastic neutron scattering experiments were performed using a powder sample to investigate the exchange interactions of this material. The spin excitations from the singlet ground state to the first three triplet states, predicted from the antiferromagnetic hexagonal spin cluster interacting with 3.9 meV, were observed. Using the exact diagonalization mothods, the intercluster coupling was estimated from the excitation peak width to be about 20\% of the intracluster interaction, which is consistent with the previously reported value.
Our exchange path model explains the anisotropic exchange interactions in the distorted honeycomb plane.

\end{abstract}

\maketitle

\section {Introduction}
A honeycomb lattice magnet is the two dimensional spin system with the minimum three exchange bonds at each spin. When the nearest-neighbor antiferromagnetic coupling is dominant, there exist no frustrating interactions and a collinear long-range antiferromagnetic order is expected as the ground state. However, including the second nearest neighbor antiferromagnetic interactions, the long-range magnetic order is suppressed and the ground state finally becomes disordered~\cite{Takano2006,Li2012}.
Another disordered ground state is proposed by Kitaev \cite{KITAEV2006} in the honeycomb lattice antiferromagnet with bond dependent exchange interactions $J_x$, $J_y$, and $J_z$. Some candidate materials realizing the Kitaev model have been studied intensively~\cite{Takagi2019}. 

\begin{figure}[b]
\includegraphics[width=8.2cm]{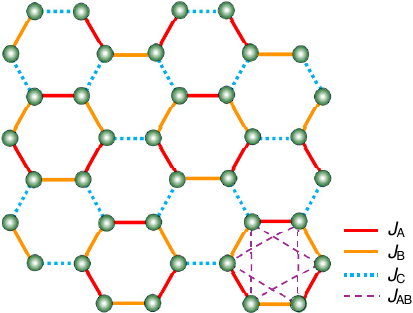}
\caption{(Color online) Schematic structures of the honeycomb lattice of Cu in Cu$_2$(pymca)$_3$(ClO$_4$). Four exchange interactions ($J\rm_{A}$, $J\rm_{B}$, $J\rm_{C}$, and $J\rm_{AB}$) are shown.
}
\label{model}
\end{figure}
Cu$_2$(pymca)$_3$(ClO$_4$), where pymca represents pyrimidine-2-carboxylate, consists of slightly distorted honeycomb lattices~\cite{Sugawara2017}. There exist three different exchange interactions, $J\rm_{A}$, $J\rm_{B}$, and $J\rm_{C}$ shown in Fig. \ref{model}, between the neighboring Cu$^{2+}$ spins due to the structural distortion in the honeycomb plane with three different bond lengths 5.5083 \AA, 5.5112 \AA, and 5.5932 \AA\ between neighboring Cu$^{2+}$ spins~\cite{Sugawara2017}. No long-range magnetic order appears down to 0.6 K. The magnetic susceptibility shows a gap behavior. However, it is not certain that the spin disordered states described above explain the magnetization, which shows 1/3 and 2/3 plateau phases~\cite{Okutani2019}.
The magnetization can rather be explained by a hexagonal spin cluster model~\cite{KOUZOUDIS1998}, in which the three neighboring exchange interactions at each Cu$^{2+}$ spin are not isotropic, and the 1/3 and 2/3 magnetization plateau phases corresponds to one-triplet and two-triplet states emerging from a hexagonal singlet state formed by coupling of three singlet states, respectively. The magnetization of Cu$_2$(pymca)$_3$(ClO$_4$) was well reproduced by the weakly coupled hexamers model~\cite{Adhikary2021,Shimokawa2022}.
Shimokawa $et$ $al.$ studied this material theoretically using the quantum Monte Carlo method and calculate the dispersion relations for the quantum honeycomb lattice system between the two extremes, $J\rm_{C}$ $\gg$ $J\rm_{A}$ = $J\rm_{B}$ and $J\rm_{A}$ = $J\rm_{B}$ $\gg$ $J\rm_{C}$~\cite{Shimokawa2022}. The former and later correspond to the weakly coupled dimers and hexamers models, respectively.

The hexagonal spin cluster system has been investigated both theoretically and experimentally.
Haraldsen performed an exact diagonalization study of the $S$ = 1/2 isotopic antiferromagnetic spin ring system, including the hexagonal spin cluster~\cite{Haraldsen2016}. The energy levels and structure factors for the ground and excited states were calculated.
The experimental realization of the magnetic excitations from antiferromagnetic hexagonal spin clusters was reported in frustrated spinel antiferromagnetic systems, consisting of the pyrochlore lattice, such as MgCr$_2$O$_4$~\cite{Tomiyasu2008,Tomiyasu2013} and HgCr$_2$O$_4$~\cite{Tomiyasu2011}, in which the Cr$^{3+}$ ion carries $S$ = 3/2. These materials exhibits flat and discrete excitations from various kinds of magnetic molecules. The first excited state can be explained with the hexagonal spin clusters. The excitations are induced due to strongly frustrated interactions and stable only in a narrow energy range.
{\color{black}Cu$_3$WO$_6$ consists of hexagonal clusters of Cu$^{2+}$ spins, which carry $S$ = 1/2. The magnetic susceptibility shows a gap behavior~\cite{Hase1995}. Inelastic neutron scattering experiment was performed to clarify the mechanism of the spin gap behavior~\cite{Hase1996}. The observed magnetic excitation data indicated that large second and third neighbor exchange interactions, which are larger than a half of the first neighbor interaction, need to be included. 
Therefore, there has been no ideal hexagonal spin cluster systems with simple exchange interactions, which have dominant nearest-neighbor exchange interaction and can be compared directly to the theories~\cite{Shimokawa2022,Haraldsen2016}.

The most interesting point in Cu$_2$(pymca)$_3$(ClO$_4$) is how the honeycomb lattice exhibits anisotropic exchange interactions, which realizes the coupled hexagonal spin clusters.}
We performed inelastic neutron scattering experiments on a powder sample of Cu$_2$(pymca)$_3$(ClO$_4$) to evaluate the exchange interactions between the Cu$^{2+}$ spins. It is not straightforward to synthesize deuterated samples of this material for neutron scattering experiments. Therefore, we aimed to observe magnetic signal using an undeuterated sample. In order to exclude the phonon signal mostly from hydrogen atoms and specify the intrinsic magnetic signal, detailed temperature dependence of the excitations is studied. Furthermore, neutron polarization analysis was performed to separate the magnetic and phonon signals. We successfully observed three triplet excited states from the singlet ground state. The observed excitations were reproduced reasonably well with the weakly coupled antiferromagnetic hexagonal spin clusters model.
{\color{black}We then considered a possible mechanism for the anisotropic exchange interactions in the slightly distorted honeycomb lattice, which explains the experimental results.}

\section {Experimental Details}
A polycrystalline sample of Cu$_2$(pymca)$_3$(ClO$_4$) was prepared by the hydrothermal reaction according to the method described in Ref. ~\onlinecite{honda2015}. The polycrystalline sample that weighs $\sim$5 g was used for the inelastic neutron scattering measurements.

The unpolarized inelastic neutron scattering experiments were carried out on the chopper neutron spectrometer SEQUOIA~\cite{SEQUOIA}, installed at the Spallation Neutron Source at Oak Ridge National Laboratory (ORNL). We utilized an incident neutron energy ($E\rm_{i}$) of 15 meV. The energy resolution at the elastic position was 0.44 meV. The measurements were performed in a temperature range of 5 $\le$ $T$ $\le$ 100 K using a closed-cycle refrigerator (CCR). The visualization of the SEQUOIA data were performed using DAVE software package~\cite{DAVE}.

Polarized inelastic neutron scattering measurements were performed on the thermal polarized triple-axis spectrometer HB-1 installed at the High Flux Isotope Reactor at ORNL. Heusler monochromator and analyzer were used with a fixed final neutron energy of 13.5 meV. The flipping ratio was about 10 estimated from nuclear Bragg peaks. The horizontal collimator sequence was 48’-80’-sample-60’-open. The energy resolution at the elastic position was 1.6 meV. The contamination from higher-order beams was eliminated using Pyrolytic Graphite (PG) filters. A CCR was used to cool down the sample to 5 K.

\section {Results and Discussion}
\begin{figure}
\includegraphics[width=8.6cm]{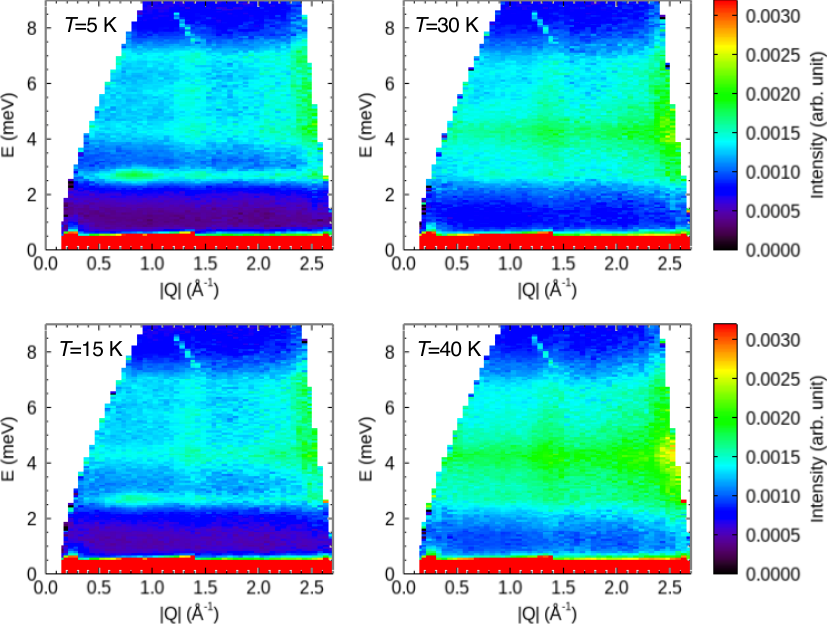}
\caption{(Color online) Contour maps of the inelastic neutron scattering intensity $S(|{\bm Q}|,E)$ from Cu$_2$(pymca)$_3$(ClO$_4$) powder measured on SEQUOIA with $E\rm_i$ = 15 meV at $T$ = 5, 15, 30, and 40 K. The background intensities, measured with an empty cell at each temperature, were subtracted. A line shaped signal around 8 meV and 1.4 \AA$^{-1}$ is spurious, originating from the instrumental configuration.}
\label{INS_exp}
\end{figure}
We first performed inelastic neutron scattering experiments using unpolarized neutrons.
Figure \ref{INS_exp} shows the excitation spectra in a powder sample of Cu$_2$(pymca)$_3$(ClO$_4$) measured on SEQUOIA as a function of temperature. There are dispersionless excitations around 2.7 meV, 4.5 meV, 5.5 meV, and 7.0 meV at 5 K. The 2.7 meV and 4.5 meV peaks become weaker and stronger at higher temperatures, respectively. This suggests that the former and latter are magnetic and phonon excitations, respectively. At higher temperatures, the phonon intensities become stronger and therefore the weak magnetic signal is difficult to distinguish.

\begin{figure}
\includegraphics[width=7.7cm]{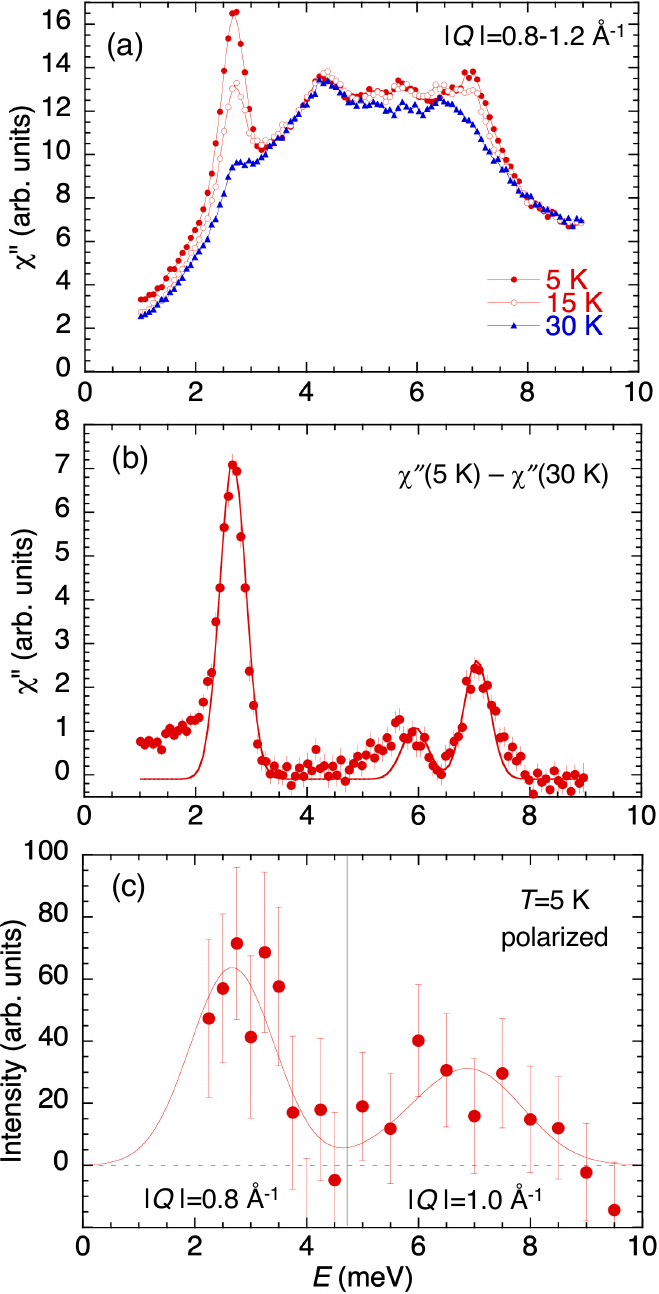}
\caption{(Color online) (a) Energy dependence of of $\chi''$ integrated with respect of $|{\bm Q}|$ in the range of 0.8 $\le |{\bm Q}| \le$ 1.2 \AA$^{-1}$ at $T$ = 5, 15, and 30 K. (b) A difference plot between 5 and 30 K. The solid line represents three Gaussian functions peaked at 2.67 meV, 5.94 meV, and 7.03 meV, predicted from the hexagonal spin cluster model with $J$ = 3.9 meV. The peak width is fixed at 0.55 meV.
(c) Difference plot between $P_{-xx}$ and $P_{xx}$ polarized channels measured on HB-1 with $E\rm_f$ = 13.5 meV at $T$ = 5 K. The solid lines are three Gaussian functions peaked at 2.67 meV, 5.94 meV, and 7.03 meV with peak widths calculated using the instrument configuration. The excitations were measured at 0.8 \AA$^{-1}$ and 1.0 \AA$^{-1}$ for the data at $E \le$ 4.5 meV and $E \ge$ 5.0 meV, respectively.}
\label{I_E}
\end{figure}

In order to subtract the nonmagnetic background intensity properly, the intensity $S(|{\bm Q}|,E)$ was converted to the dynamic susceptibility $\chi''(|{\bm Q}|,E)$ using the relation, $\chi''(|{\bm Q}|,E)$ = $S(|{\bm Q}|,E)\cdot$[1$-$exp($-E/k{\rm_B}T$)], where ${\bm Q}$, $E$, $k\rm_B$, and $T$, are scattering wave vector, excitation energy, the Boltzmann constant, and temperature, respectively. Figure \ref{I_E}(a) exhibits $\chi''(E)$ integrated with respect of $|{\bm Q}|$ in the range of 0.8 $\le |{\bm Q}| \le$ 1.2 \AA$^{-1}$ at $T$ = 5, 15, and 30 K. Since the thermal factor is corrected, the phonon component shows no temperature dependence. At higher temperatures, the phonon component starts to deviate slightly probably due to anharmonicity (not shown). The magnetic excitation peak at $\sim$2.7 meV becomes reduced with increasing temperature because the population at the ground state becomes smaller, which makes the transition between the two states less. This plot also shows the decrease of intensity around 5.5 meV and 7 meV. Figure \ref{I_E}(b) shows the difference of $\chi''(E)$ at 5 K and 30 K. This clearly shows excitation peaks around 2.7, 5.5, and 7.0 meV, which are considered to be magnetic in origin.

An additional polarized neutron scattering experiment was performed on HB-1 to confirm that the excitation peaks are magnetic. The same sample used for the unpolarized neutron scattering experiment was measured. By measuring the spin-flip ($P_{-xx}$) and non-spin-flip channels ($P_{xx}$) with aligning the neutron spin direction along ${\bm Q}$, the magnetic and phonon components can be separated. Figure \ref{I_E}(c) is the plot of the difference between the two channels, which corresponds to the magnetic component. Although the intrinsic magnetic signal is weak and the energy resolution is broad, the result supports that the three peaks in Fig. \ref{I_E}(b) are magnetic.

As mentioned in Sec. I, the magnetization study suggested that weakly coupled hexagonal clusters model with small $J\rm_C$ is appropriate to describe the spin system. Therefore, the result was analyzed using the hexagonal spin cluster model, which corresponds to $J\rm_{A}$ = $J\rm_{B}$ and $J\rm_{C}$ = 0 in Fig. \ref{model}. The isolated antiferromagnetic hexagonal spin cluster model with the intracluster interaction $J$ gives rise to triplet excited states at 0.685$J$, 1.522$J$, 1.803$J$, 2.921$J$, 3.303$J$, and 3.584$J$~\cite{Haraldsen2016}. With $J$(= $J\rm_A$ = $J\rm_B$) = 3.9 meV, the first three triplet excited states are expected at 2.67, 5.94, and 7.03 meV. The solid curve in Fig. \ref{I_E}(b) is the Gaussian peaks at these calculated positions with the peak width of 0.55 meV. Although the observed values below 2 meV and between 5 and 5.5 meV are not reproduced well, the overall agreement is reasonably well.
The exchange interaction of $J$(= 3.9 meV) is very close to 43.7 K(= 3.8 meV) obtained from the magnetization results~\cite{Okutani2019}.
If the next-nearest-neighbor interaction $J\rm_{AB}$ is finite, the second and third excited states should shift to lower energies~\cite{Haraldsen2016}. Since the excitation to the third triplet state was observed at the energy  expected from the hexagonal spin cluster model without $J\rm_{AB}$, $J\rm_{AB}$ should be negligibly small.

\begin{figure}
\includegraphics[width=7.7cm]{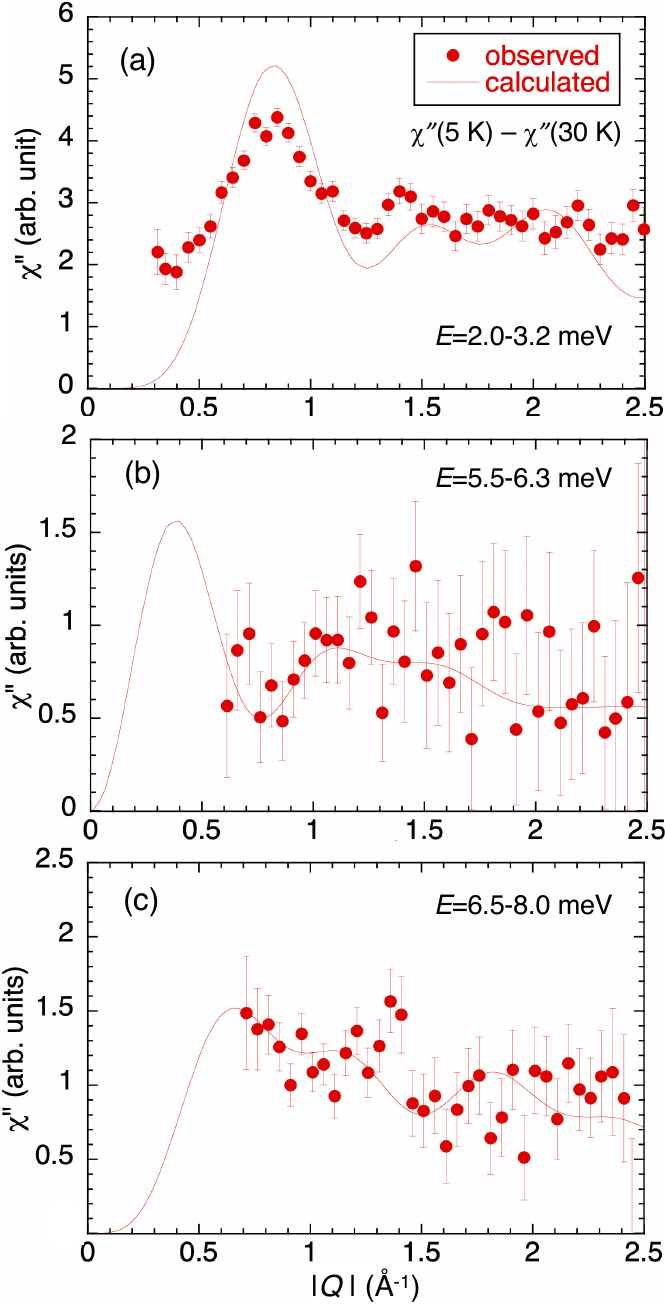}
\caption{(Color online) Momentum transfer ($|{\bm Q}|$) dependence of $\chi''$ integrated with respect of $E$ in the range of 2.0 $\le E \le$ 3.2 meV (a), 5.5 $\le E \le$ 6.3 meV (b), and 6.5 $\le E \le$ 8.0 meV (c) at $T$ = 5 K. The background intensities at 30 K were subtracted. The solid lines represent $\chi''(|{\bm Q}|)$ predicted from the hexagonal spin cluster model~\cite{Haraldsen2016}. The squared magnetic form factor for the Cu$^{2+}$ spin is included in the calculation.}
\label{I_Q}
\end{figure}
The structure factor of each excited state as a function of $|{\bm Q}|$ was analyzed to evaluate the hexagonal spin cluster model. Figures \ref{I_Q}(a), (b), and (c) display the $|{\bm Q}|$ dependence of $\chi''$ around 2.7, 5.9 and 7.0 meV, respectively. As in Fig. \ref{I_E}(b), the $\chi''$ data measured at 30 K was subtracted as background. Since the difference in $\chi''$ is small at 5.9 and 7.0 meV, the error bars are large. The solid line in each figure represents the calculated value for the antiferromagnetic hexagonal spin cluster model. At 2.7 meV, the first peak position at $\sim$0.8 \AA$^{-1}$ is reproduced well. Due to possible phonon scattering around 1.4 \AA$^{-1}$, as shown in Fig. \ref{INS_exp}, the background subtraction may not be done perfectly so that a bump is seen around the $|{\bm Q}|$ position. Except $|{\bm Q}| \sim$ 1.4 \AA$^{-1}$, the overall agreement is reasonably well.

The hexagonal spin cluster model predicts ten magnetic excited states in total~\cite{Haraldsen2016}. Therefore, seven more excited states are expected at higher energies. We tried to detect the fourth excited state expected at 11.4 meV. However, the intensity was too weak to be observed, consistent with the theoretical prediction~\cite{Shimokawa2022}. Since higher energy excitations are expected to be even weaker, it would be difficult to observe the excited states other than the first three experimentally in this material.

To investigate the effect of the $J_{\rm C}$ interaction on the width of the excitation peaks and obtain a good estimation of the $J_{\rm C}$, we calculate the integrated dynamical spin structure factor based on exact diagonalization methods. 
Let us briefly introduce the numerical procedures. As in the previous paper~\cite{Shimokawa2022}, we here define our spin Hamiltonian
\begin{eqnarray}
{\hat \mathcal{H}}=J_{\rm A}\sum_{i,j} {\bm S}_i \cdot {\bm S}_j + J_{\rm B}\sum_{i,j} {\bm S}_i \cdot {\bm S}_j + J_{\rm C}\sum_{i,j} \textbf{\emph S}_i \cdot \textbf{\emph S}_j,
\end{eqnarray}
and use a 36-site finite-size cluster, consisting of six hexamers. We consider 4 parameter cases of $J_{\rm A}/J_{\rm C}=J_{\rm B}/J_{\rm C}=$ 5.0, 10.0, 15.0 and 20.0 for our computations.

We first compute zero-temperature $z$-component of the dynamical spin structure factor at each accessible wave vector ${\bm Q}$ on the 36-site finite-size cluster with using the following equation,
\begin{eqnarray}
S_{\rm L}({\bm Q}, E)= -\frac{3}{\pi} {\rm Im} \langle \phi| {\hat S}_{\bm Q}^{z \dagger}  \frac{1}{E - {\hat \mathcal{H}} + E_0 + i\eta} {\hat S}_{\bm Q}^{z } | \phi \rangle,
\end{eqnarray}
where the $|\phi \rangle$ and $E_0$ are the ground-state wave function and the corresponding energy obtained by Lanczos algorithm.
This equation is rewritten with using continued fraction~\cite{Gagliano87}  as 
\begin{eqnarray}
S_{\rm L}({\bm Q}, E) = -\frac{3}{\pi} {\rm Im} \frac{\langle \phi| \hat{S}^{z\dagger}_{\bm Q} \hat{S}^{z}_{\bm Q} |\phi \rangle} {z-\alpha_1-\frac{\beta^2_1}{z-\alpha_2-\frac{\beta^2_2}{z-\alpha_3-\cdots}}}
\end{eqnarray}
with $z$ = $E - E_0 + i\eta$. $\alpha$ and $\beta$ are obtained by the tridiagonalization procedure of the Hamiltonian matrix in the Lanczos iteration.  
The $\eta$ is the broadning factor described by Lorentzian function, and we set as $\eta$ = 0.10.

We should note that experimental results were described using a Gaussian-type dynamical spin structure factor so that once each dynamical spin structure factor is evaluated using a Lorentzian function, 
we extract the peak positions, $\{ {E_{ n} } \}$, and the corresponding peak values $\{ S_{\rm L}({\bm Q}, E_{n}) \}$  and use each of them to fit a Gaussian function.
Therefore, we could obtain the Gaussian-type dynamical spin structure factor as
\begin{eqnarray}
S_{\rm G}({\bm Q}, E)= \sum_{n=1}^{3} S_{\rm L}({\bm Q}, E_{n}) \exp(-\frac{(E-E_n)^2}{2 \alpha^2})
\end{eqnarray}
The constant value $\alpha$ is known as standard deviation of the Gaussian function, and this is related to the instrumental energy resolution $\Delta E_{\rm res}$, which is the full width at the half maximum (FWHM).
The relationship between the $\alpha$ and $\Delta E_{\rm res}$ is written by 
\begin{eqnarray}
\Delta E_{\rm res}=2 \alpha \sqrt{2 {\rm ln}(2)},
\end{eqnarray}
and we set $\alpha$ based on $\Delta E_{\rm res}$ = 0.35 meV, which corresponds to the instrumental energy resolution for the lowest energy peak around 2.7 meV.
After evaluating $S_{\rm G} ({\bm Q}, E)$ based on the above procedures, we integrate  $S_{\rm G} (|{\bm Q}|, E)$ over the range 0.8 $\le$ $|{\bm Q}|$ $\le$ 1.2 ${\rm \AA^{-1}}$ by summing over the discrete wave vectors ${\bm Q}$ that fall within the range.

\begin{figure}[t]
  \centering
  \subfloat[]{
    \includegraphics[width=0.8\columnwidth]{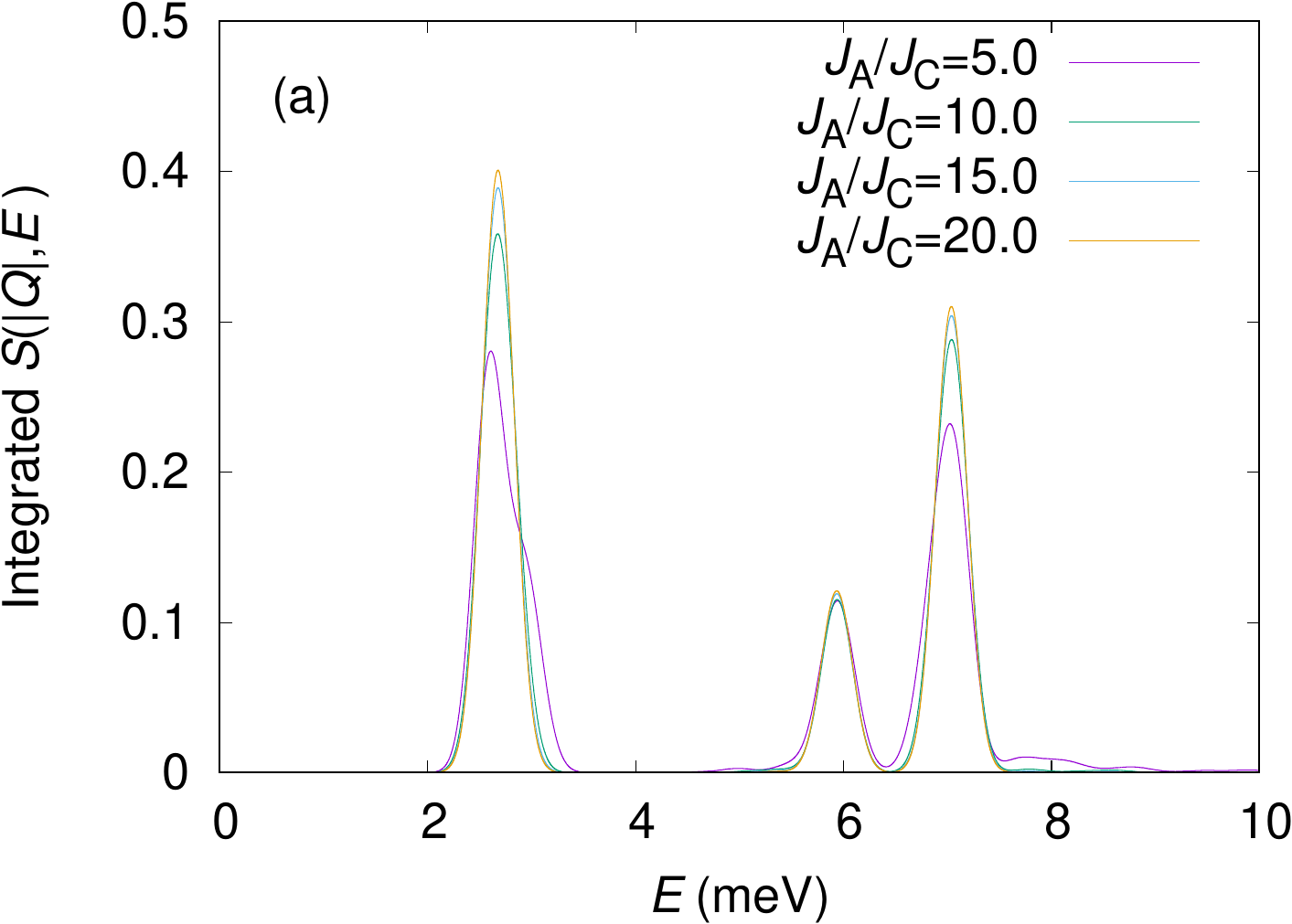}
    \label{fig:ISQW}
  }
  \vspace{-7mm}
  \\
  \subfloat[]{
    \includegraphics[width=0.8\columnwidth]{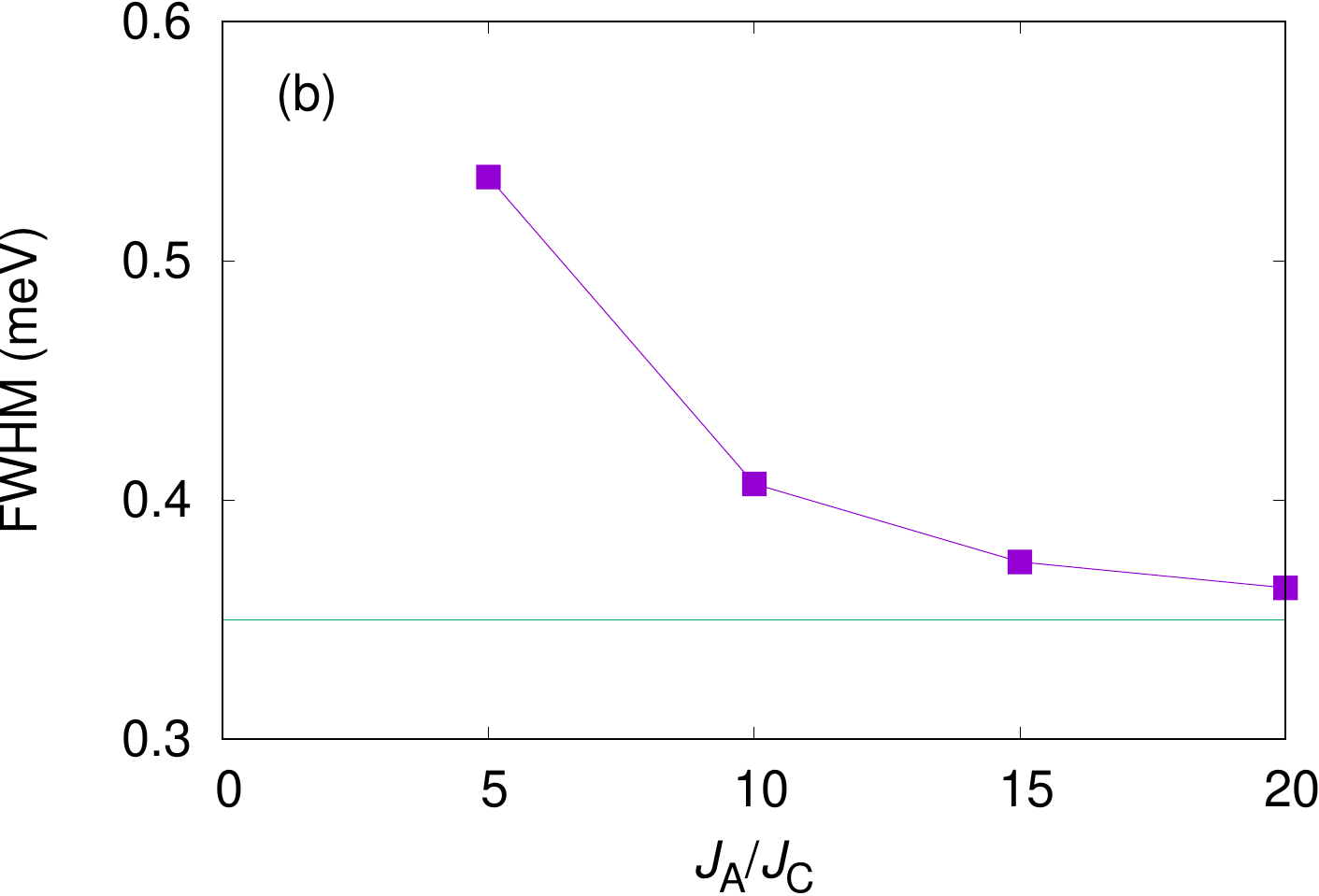}
    \label{fig:FWHM}
  }
  \caption{Exact diagonalization results of the integrated dynamical spin structure factor for $N$ = 36 spin cluster (six hexamers). (a) The $J_{\rm A}/J_{\rm C}$ dependence of the $S_{\rm G} (|{\bm Q}|, E)$ integrated over 0.8 $\le$ $|{\bm Q}|$ $\le$ 1.2 ${\rm \AA^{-1}}$, where the $S_{\rm G} (|{\bm Q}|, E)$ is the Gaussian-type dynamical spin structure factor (see main text). The instrumental energy resolution is convoluted. (b) The evaluated full width at half maximum (FWHM) of the peak around 2.7 meV in (a). The horizontal line indicates the instrumental energy resolution for the peak.}
  \label{fig:ED}
\end{figure}

The computational results of the integrated spin dynamics are shown in Fig.~\ref{fig:ED}(a). The energies of the first three excited peaks are consistent with those predicted from the hexagonal cluster model~\cite{Haraldsen2016}.
With increasing $J_{\rm C}$, each excitation peak becomes more dispersive~\cite{Shimokawa2022}, which makes the peak broader.
{\color{black}
It is noted that the heights of the three excitation peaks are comparable, which is different from the experimental result shown in Fig. \ref{I_E}(b). The primary cause of this disrepancy may be that the background is oversubtracted for the two higher energy excitations, where the background is much larger than the intrinsic signal.}
The $J_{\rm A}/J_{\rm C}$ dependence of FWHM for the peak around 2.7 meV is shown in Fig.~\ref{fig:ED}(b). With reducing $J_{\rm C}$, the observed peak width becomes narrower and approach to the instrumental energy resolution.
Since the peak width was observed most accurately for the peak at the lowest energy and was predicted to be strongly dependent of $J_{\rm C}$ theoretically, we estimate $J_{\rm C}$ using the width of this peak. As shown in Fig.~\ref{fig:ED}(b), the observed peak width (0.55 meV) can be reproduced well with $J_{\rm A}$/$J_{\rm C}$ = 5.0.
This is consistent with the previously reported value of $J_{\rm A}$/$J_{\rm C}$ = 5.0, which is obtained from the magnetization and magnetic susceptibility measurements~\cite{Okutani2019,Shimokawa2022}.

\begin{figure}
\includegraphics[width=8.6cm]{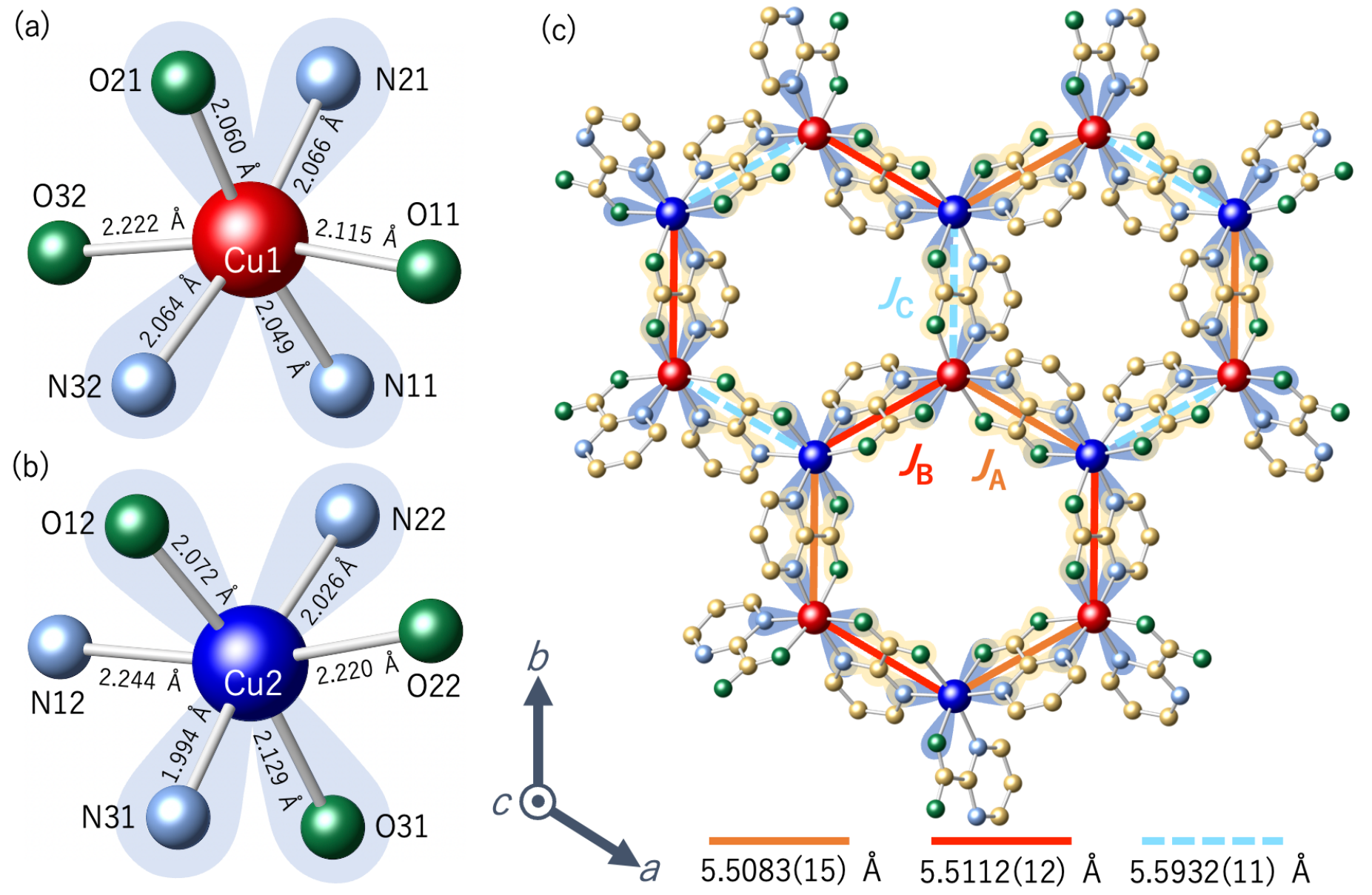}
\caption{(Color online) Local environments around Cu1 (a) and Cu2 (b) ions with $d_{x^2-y^2}$ orbital. (c) Crystal structure and exchange paths between the neighboring Cu$^{2+}$ spins in the honeycomb plane of Cu$_2$(pymca)$_3$(ClO$_4$). There are three different pymca (C$_5$N$_2$O$_2$) units bridging the Cu$^{2+}$ spins. The shaded regions represent the $d_{x^2-y^2}$ orbitals that mediate the exchange bonds.}
\label{exchange}
\end{figure}
As mentioned in Sec. I, bond lengths for the three neighboring spins at each Cu spin are different (5.5083 \AA, 5.5112 \AA, and 5.5932 \AA)~\cite{Sugawara2017}. However, from the present results, we cannot specify which bond corresponds to $J\rm_{A}$ or $J\rm_{B}$ or $J\rm_{C}$ experimentally. Even with single crystal experiments, it would be extremely difficult to specify the interactions due to the tiny difference between the three bond lengths.
Here, the exchange interactions between the Cu$^{2+}$ spins in the honeycomb plane are discussed using the structural data reported in Ref. \onlinecite{Sugawara2017}. As shown in Figs. \ref{exchange}(a) and \ref{exchange}(b), each Cu$^{2+}$ ion has six bonds with three Cu-O and three Cu-N bonds. Four bonds are shorter than the other two bonds. Therefore, the exchange interactions are considered to be mediated via the $d_{x^2-y^2}$ orbitals along the four shorter bonds{\color{black}, as shown with shaded regions in Figs. \ref{exchange}(a) and \ref{exchange}(b). The $sp^{\rm 2}$ orbitals on the carbon atoms between Cu1 and Cu2 are also shown.
There are three different pymca (C$_5$N$_2$O$_2$) units bridging the {\color{black}Cu1 and Cu2} ions.
Since the two pymca units out of three are structurally similar, we assume that these corresponds to the bonds for $J\rm_A$ and $J\rm_B$, as shown in Fig. \ref{exchange}(c). The orbital overlap of the Cu-N-C-N-Cu exchange path in $J\rm_A$ and $J\rm_B$ is effective, which leads to strong exchange interactions.
$J\rm_A$ and $J\rm_B$ also have the Cu-N-C-C-O-Cu exchange path, although the orbital overlap is not as good as that of the Cu-N-C-N-Cu exchange path.
On the other hand, the pymca unit for $J\rm_C$ only has the Cu-N-C-C-O-Cu exchange path. Therefore, the three exchange interactions ($J\rm_A$, $J\rm_B$, and $J\rm_C$) are not equivalent but one exchange bond has a reduced exchange interaction. This is consistent with the experimental results, which indicates that $J\rm_A$ and $J\rm_B$ are much larger than $J\rm_{C}$.}

As discussed in Refs. \onlinecite{Okutani2019,Shimokawa2022}, the magnetization shows a bending between the 2/3 plateau and full saturation state. This might be due to additional further neighbor interactions or magnetic anisotropies, such as the Dzyaloshinskii–Moriya interaction.
The present results suggest that these interactions are not distinct enough to affect the excitation energies or peak widths.
Further theoretical studies are required to solve this issue.

\section {Summary}
In summary, inelastic neutron scattering experiments were performed on a powder sample of the distorted honeycomb lattice Cu$_2$(pymca)$_3$(ClO$_4$) to study the exchange interactions between Cu$^{2+}$ spins. We successfully separated three triplet excited states. The magnetic excitations are reproduced reasonably well with the weakly coupled antiferromagnetic hexagonal spin clusters model, which is explained using our exchange path model. Interestingly, Cu$_2$(pymca)$_3$(ClO$_4$) is found to be a rare example for the quantum hexagonal spin cluster system.

\section *{Acknowledgments}
This research used resources at the Spallation Neutron Source and High Flux Isotope Reactor, DOE Office of Science User Facilities operated by the Oak Ridge National Laboratory. The beam time was allocated to SEQUOIA and HB-1 on proposal numbers IPTS-33649.1 and IPTS-34804.1, respectively.
T. S. is supported by the Theory of Quantum Matter Unit of the 
Okinawa Institute of Science and Technology Graduate University (OIST), 
and this work is also supported by JSPS Grant-in-Aid for Scientific Research (C) 
Grants No. 21K03477 and 25K07213, and MEXT Grant-in-Aid for Transformative 
Research Areas A ``Extreme Universe'' Grant No. 22H05266 and 24H00974.
Exact diagonalization calculations were carried out using HPC facilities provided by  
the Supercomputing Center, ISSP, the University of Tokyo and OIST.

\bibliography{hexamer.bib}

\end{document}